\documentstyle[conf_iap]{article}

\begin{document}

\heading{Galaxy as dissipative N -- body system. \\
         Chemical aspect.} 

\par\medskip\noindent

\author{Peter Berczik, Sergei G. Kravchuk \bigskip} 
       {Main Astronomical Observatory of Ukrainian National 
       Academy of Sciences \\
       252650, Golosiiv, Kiev-022, Ukraine, e-mail: berczik@mao.kiev.ua \bigskip}


\noindent {\bf Abstract.}  The evolving galaxy is considered as a
system of baryonic fragments embedded into the static dark
nonbaryonic (DH) and baryonic (BH) halo and subjected to
gravitational and viscous interactions. Though the chemical
evolution of each fragment is treated in the frame of one -- zone
model with instantaneous recycling, the local gas density and,
therefore, its star formation (SF) activity is influenced by other
interacting fragments. In spite of its simplicity the model provides
a realistic description of the process of galaxy formation and
evolution over the Hubble timescale.


\section{The model}

     The Plummer -- type (PLT) \cite{DC95} halo is composed of DH and BH 
components. Their masses and softening parameters are correspondently 
equal to $ 10^{12} \; M_\odot $, $ 10^{11} \; M_\odot $, and $ 25 \; kpc 
$, $ 15 \; kpc $. The clumped baryonic matter (future galaxy disk and 
bulge) of mass $ 10^{11}\; M_\odot $ is equally distributed among $ N = 
2109 $ PLT particles with softening parameters $ 1 \; kpc $. Initially 
particles are smoothly placed inside the sphere of radius $ 50 \; kpc $, 
which rigidly rotate around $ z $ axies with  initial angular velocity 
that provide rotational support of matter in its outer equatorial 
region. They are involved into the Hubble flow ($ H_{0} = 75 \; km/s 
\cdot Mpc^{-1} $) and have random velocities $ 0 \div 10 \; km/sec $. 
Their subsequent dynamics is defined by particle -- halo and particle -- 
particle gravitational interactions and by viscous BH influence and is 
followed by means of effective N -- body integrator with individual time 
step. Gravitational terms are defined in a standard manner. The 
comparison of model results with data of more sophisticated SPH 
approaches shows that viscous acceleration term can be choosed as $ {\bf 
A}^{VISC}_i = \\ - {\bf V}_i \cdot ~\mid {\bf V}_i \mid/R_{VISC} \cdot 
\rho_{BH}(r)/\rho_{VISC} \cdot m^{gas}_i/(m^{gas}_i +m^{star}_i) $. Here 
$ \rho_{BH} $ is the baryonic halo local density and $ {\bf V}_i $ is a 
particle velocity vector. $ \rho_{VISC} = 0.2 \; cm^{-3} $ and $ 
R_{VISC} = 10 \; kpc $ are model parameters. A particle mass is defined 
as a sum of gas $ m^{gas}_i $ and stellar $ m^{star}_i $ components 
which depend on its evolutionary status. Initially $ m^{star}_i = 0 $. 
To account for the influence of baryonic halo rotation onto the dynamics 
of baryonic particles in the disk galaxy plane $ {\bf A}^{VISC} $ is set 
equal to $ (0.15,0.15, 1.0) \cdot {\bf A}^{VISC} $.

      The characteristic time step in the integration procedure for
a particle is $ \delta t_i = 10^{-2} \cdot \min_{j} [ \sqrt{ \mid
{\bf R}_{ij} \mid / \mid {\bf A}_{ij} \mid }, \mid {\bf R}_{ij} \mid
/ \mid {\bf V}_{ij} \mid ] $, where the $ {\bf V}_{ij} = {\bf V}_i -
{\bf V}_j $ and the $ {\bf A}_{ij} = {\bf A}_i - {\bf A}_j $ is a
difference of accelerations of $ i $ and $ j $ particles. Such step
value provides a nice momentum and energy conservation over the
integration interval of about $ 15 \cdot 10^9 \; year $.

     The local gas density $ \rho_i $ is defined in a SPH -- like
manner as a ratio of mass of gas contained in the sphere of radius $
H_i $ around the $ i $ -- th particle to its volume. The $ H_i $ is
defined (using the quicksort algorithm) by the requirement that the
number of particles inside this sphere is equal to $ N_B = 21 $ ($
\approx $ 1 \% of $ N $). The particle eligible to star formation
(SF) events is choosed as particle which still have a sufficient
amount of gas (exceeding $ 0.01 $ of particle initial mass) and
densities exceeding the critical value $ \rho_{minSF} = 0.01 \;
cm^{-3} $ during the fixed time interval $ \Delta t_{SF} = 50 \;
Myear $. Such active particle produces stars with a Scalo -- type
initial mass function with SF efficiency $ \epsilon $ is defined as
$ \epsilon = 0.5 \cdot \rho_i / 10 \; cm^{-3} \cdot (1 - exp(- 10 \;
cm^{-3} / \rho_i )) $. Each SF event is followed by SN explosions
which inject into the particle the enriched gas of metallicity $ Z_i
+  \Delta Z $. Initially $ Z_i = 0.0 $ and $ \Delta Z = 0.01 $ as the
average value of enrichment for stars having metallicities in the
range $ Z = 0.001 \div 0.04 $. The returned fraction of gas is set
equal to $ R = 0.25 $. This gas is assumed to be immediately mixed
with old (heavy element deficient) gas.


\section{Conclusion}

     The rapidly rotating protogalaxy formes a three - component
system resembling a typical spiral galaxy: a thin disk and
spheroidal component made of gas and stars and a dark matter halo.
The disk component poseses a typical spiral galaxy rotation curve
and the distribution of radial and z - th velocities of particles
clearly shows the presence of the central bulge. The metallicities
and the global metallicity gradient resemble distributions observed
in spiral galaxies. During the first 2 Gyr of evolution only about $
20 \% $ of protogalaxy baryonic component is transformed into the
stars, but, finally, the  fraction of the stellar population grows
up to $ 92 \% $. The galaxy global star formation rate (SFR) can be
considered as a continuous succesion of short bursts which initially
doesn't exceed $ 28 \; M_{\odot} \; yr^{-1} $. Then SFR gradually
decreases up to the value of about $ 2 \; M_{\odot} \; yr^{-1} $ .
All these data as well as surface densities distributions of stellar
and gaseous components are in nice agreement with present date
observational data of our own Galaxy \cite{KG89, P94} and with
results of more sophisticated theoretical approaches
\cite{SM94, RVN96, SH96, BK97}.


\baselineskip=0.1cm

\acknowledgements{Peter Berczik would like to acknowledge the Open
Society Institute for partial financial support of this work under
grants No. {\bf TE7 100} and is especially thanks to Organizing
Committee for financial support.}



\begin{thebibliography}{99}

\bibitem{BK97} Berczik P., Kravchuk S.G., 1997, Ap.Sp.Sci. {\bf 245}, 27
\bibitem{DC95} Douphole B., Colin J., 1995, A.\&A. {\bf 300}, 117
\bibitem{KG89} Kuijken K. Gilmore G., 1989, M.N.R.A.S. {\bf 239}, 605
\bibitem{P94} Pagel B.E.J., 1994, in "The Formation and Evolution of Galaxies", Eds. Munoz - Tunon, C., Sanchez, F., Cambridge Univ. Press
\bibitem{RVN96} Raiteri C.M., Villata M., Navarro J.F., 1996, A.\&A. {\bf 315}, 105
\bibitem{SH96} Samland M., Hensler G., 1996, Reviews in Modern Astronomy {\bf 9}, 277
\bibitem{SM94} Steinmetz M., Muller E., 1994, A.\&A. {\bf 281}, L97

\end{thebibliography}
\end{document}